\documentclass{llncs}
\usepackage{epsfig}
\usepackage[utf8]{inputenc}
\usepackage{multicol}
\usepackage{multirow}
\usepackage{wrapfig}
\usepackage{fancyvrb}

\graphicspath{{./figures/}}
\newcommand{\verbatimproperties}{\renewcommand{\baselinestretch}{0.85} \small}
\newcommand{\tableproperties}{\centering \small}

\begin{document}

\title{Global Trie for Subterms}

\author{João Raimundo \and Ricardo Rocha}

\institute{CRACS \& INESC-Porto LA, Faculty of Sciences, University of Porto\\
           Rua do Campo Alegre, 1021/1055, 4169-007 Porto, Portugal\\
           \email{\{jraimundo,ricroc\}@dcc.fc.up.pt}}

\maketitle


\begin{abstract}
  A critical component in the implementation of an efficient tabling
  system is the design of the table space. The most popular and
  successful data structure for representing tables is based on a
  two-level trie data structure, where one trie level stores the
  tabled subgoal calls and the other stores the computed answers. The
  Global Trie (GT) is an alternative table space organization designed
  with the intent to reduce the tables's memory usage, namely by
  storing terms in a global trie, thus preventing repeated
  representations of the same term in different trie data
  structures. In this paper, we propose an extension to the GT
  organization, named \emph{Global Trie for Subterms (GT-ST)}, where
  compound subterms in term arguments are represented as unique
  entries in the GT. Experiments results using the YapTab tabling
  system show that GT-ST support has potential to achieve significant
  reductions on memory usage, for programs with increasing compound
  subterms in term arguments, without compromising the execution time
  for other programs.\\

\end{abstract}


\section{Introduction}

Tabling~\cite{Chen-96} is an implementation technique that overcomes
some limitations of traditional Prolog systems in dealing with
redundant sub-computations and recursion. Tabling became a renowned
technique thanks to the leading work in the XSB-Prolog system and, in
particular, in the SLG-WAM engine~\cite{Sagonas-98}. A critical
component in the implementation of an efficient tabling system is the
design of the data structures and algorithms to access and manipulate
the \emph{table space}. The most popular and successful data structure
for representing tables is based on a two-level \emph{trie data
  structure}, where one trie level stores the tabled subgoal calls and
the other stores the computed answers~\cite{RamakrishnanIV-99}.

Tries are trees in which common prefixes are represented only
once. The trie data structure provides complete discrimination for
terms and permits look up and possibly insertion to be performed in a
single pass through a term, hence resulting in a very efficient and
compact data structure for term representation. Despite the good
properties of tries, one of the major limitations of tabling, when
used in applications that pose many queries and/or have a large number
of answers, is the overload of the table space
memory~\cite{Rocha-05b}.

The \emph{Global Trie (GT)}~\cite{CostaJ-09a,CostaJ-09b} is an
alternative table space organization where tabled subgoal calls and
tabled answers are represented only once in a \emph{global trie}
instead of being spread over several different trie data
structures. The major goal of GT's design is to save memory usage by
reducing redundancy in the representation of tabled calls/answers to a
minimum.

In this paper, we propose an extension to the GT organization, named
\emph{Global Trie for Subterms (GT-ST)}, where compound subterms in
term arguments are represented as unique entries in the GT. Our new
design extends a previous design, named \emph{Global Trie for Terms
  (GT-T)}~\cite{CostaJ-09b}, where all argument and substitution
compound terms appearing, respectively, in tabled subgoal calls and
tabled answers are already represented only once in the
GT. Experiments results, using the YapTab tabling
system~\cite{Rocha-05a}, show that GT-ST support has potential to
achieve significant reductions on memory usage for programs with
increasing compound subterms in term arguments, when compared with the
GT-T design, without compromising the execution time for other
programs.

The remainder of the paper is organized as follows. First, we
introduce some background concepts about tries and the original table
space organization in YapTab. Next, we present the previous GT-T
design. Then, we introduce the new GT-ST organization and describe how
we have extended YapTab to provide engine support for it. At last, we
present some experimental results and we end by outlining some
conclusions.


\section{YapTab's Original Table Space Organization}

The basic idea behind a tabled evaluation is, in fact, quite
straightforward. The mechanism basically consists in storing, in the
table space, all the different tabled subgoal calls and answers found
when evaluating a program. The stored subgoal calls are then used to
verify if a subgoal is being called for the first time or if it is a
repeated call. Repeated calls are not re-evaluated against the program
clauses, instead they are resolved by consuming the answers already
stored in the table space. During this process, as further new answers
are found, they are stored in their tables and later returned to all
repeated calls.

The table space may thus be accessed in a number of ways: (i) to find
out if a subgoal is in the table and, if not, insert it; (ii) to
verify whether a newly found answer is already in the table and, if
not, insert it; and (iii) to load answers from the tables to the
repeated subgoals. With these requirements, a correct design of the
table space is critical to achieve an efficient implementation. YapTab
uses \emph{tries} which is regarded as a very efficient way to
implement the table space~\cite{RamakrishnanIV-99}.

A trie is a tree structure where each different path through the
\emph{trie nodes} corresponds to a term described by the tokens
labelling the nodes traversed. Two terms with common prefixes will
branch off from each other at the first distinguishing token. For
example, the tokenized form of the term $f(X,g(Y,X),Z)$ is the
sequence of 6 tokens: $f/3$, $VAR_0$, $g/2$, $VAR_1$, $VAR_0$ and
$VAR_2$, where each variable is represented as a distinct $VAR_i$
constant~\cite{Bachmair-93}. YapTab's original table design implements
tables using two levels of tries, one level stores the tabled subgoal
calls and the other stores the computed answers.

\begin{figure}[!ht]
\centering
\includegraphics[width=10.5cm]{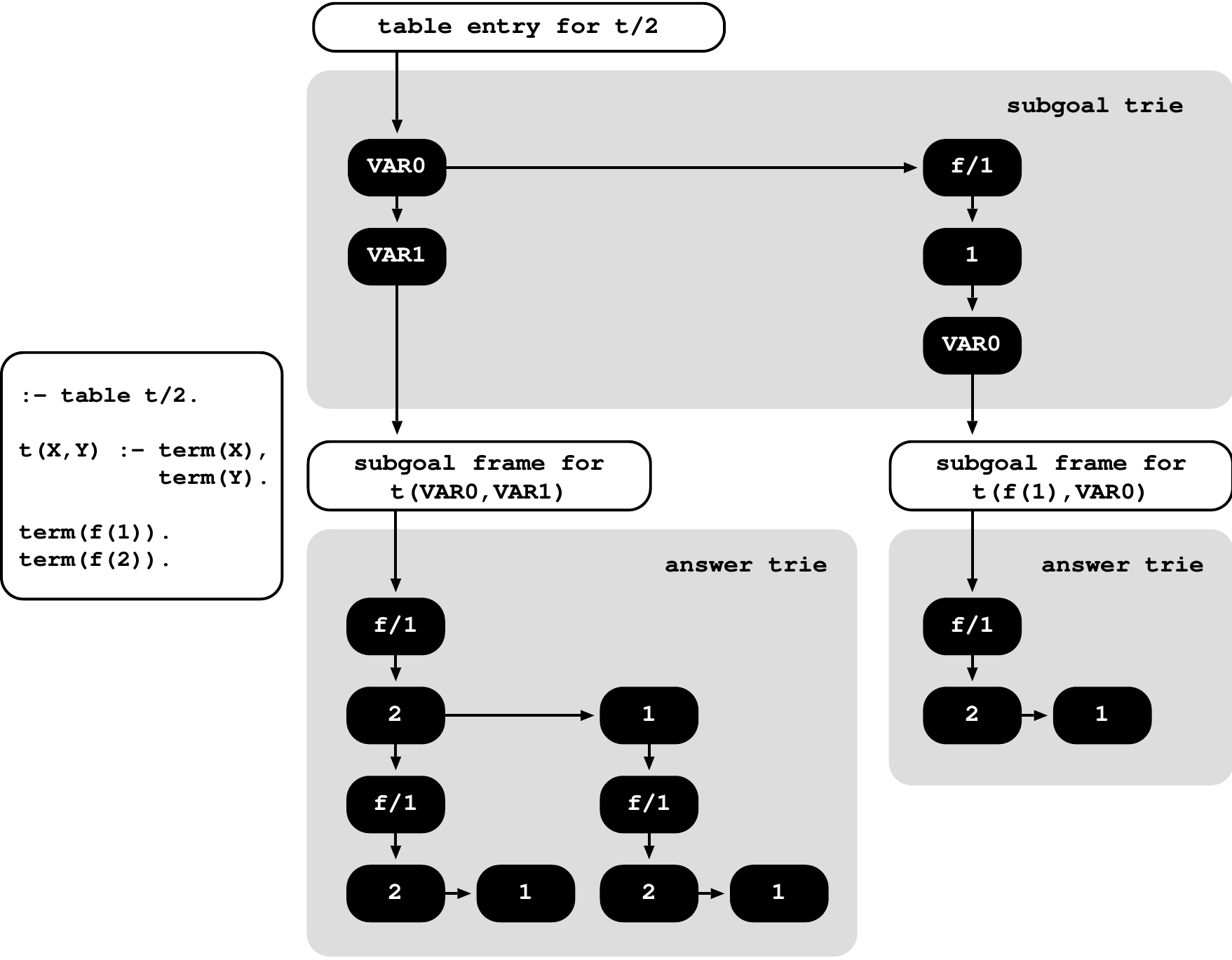}
\caption{YapTab's original table space organization}
\label{fig_original_design}
\end{figure}

More specifically, each tabled predicate has a \textit{table entry}
data structure assigned to it, acting as the entry point for the
predicate's \textit{subgoal trie}. Each different subgoal call is then
represented as a unique path in the subgoal trie, starting at the
predicate's table entry and ending in a \textit{subgoal frame} data
structure, with the argument terms being stored within the path's
nodes. The subgoal frame data structure acts as an entry point to the
\textit{answer trie}. Each different subgoal answer is then
represented as a unique path in the answer trie. Contrary to subgoal
tries, answer trie paths hold just the substitution terms for the free
variables which exist in the argument terms of the corresponding
subgoal call~\cite{RamakrishnanIV-99}. Repeated calls to tabled
subgoals load answers by traversing the answer trie nodes
bottom-up. An example for a tabled predicate \texttt{t/2} is shown in
Fig.~\ref{fig_original_design}.

Initially, the subgoal trie is empty. Then, the subgoal
\texttt{t(f(1),Y)} is called and three trie nodes are inserted: one
for functor \texttt{f/1}, a second for integer \texttt{1} and one last
for variable \texttt{Y} (\texttt{VAR0}). The subgoal frame is inserted
as a leaf, waiting for the answers. Next, the subgoal \texttt{t(X,Y)}
is also called. The two calls differ in the first argument, so tries
bring no benefit here. Two new trie nodes, for variables \texttt{X}
(\texttt{VAR0}) and \texttt{Y} (\texttt{VAR1}), and a new subgoal
frame are inserted. Then, the answers for each subgoal are stored in
the corresponding answer trie as their values are computed. Subgoal
\texttt{t(f(1),Y)} has two answers, \texttt{Y=f(1)} and
\texttt{Y=f(2)}, so we need three trie nodes to represent both: a
common node for functor \texttt{f/1} and two nodes for integers
\texttt{1} and \texttt{2}. For subgoal \texttt{t(X,Y)} we have four
answers, resulting from the combination of the answers \texttt{f(1)}
and \texttt{f(2)} for variables \texttt{X} and \texttt{Y}, which
requires nine trie nodes to represent them. Note that, for this
particular example, the completed answer trie for \texttt{t(X,Y)}
includes in its representation the completed answer trie for
\texttt{t(f(1),Y)}.


\section{Global Trie}

In this section, we introduce the new \emph{Global Trie for Subterms
  (GT-ST)} design. Our new proposal extends a previous design named
\emph{Global Trie for Terms (GT-T)}~\cite{CostaJ-09b}. We start by
briefly presenting the GT-T design and then we discuss in more detail
how we have extended and optimized it to our new GT-ST approach.


\subsection{Global Trie for Terms}

The GT-T was designed in order to maximize the sharing of tabled data
which is structurally equal. In GT-T, all argument and substitution
compound terms appearing, respectively, in tabled subgoal calls and
tabled answers are represented only once in the GT, thus preventing
situations where argument and substitution terms are represented more
than once as in the example of Fig.~\ref{fig_original_design}.

\begin{figure}[!ht]
\centering
\includegraphics[width=9cm]{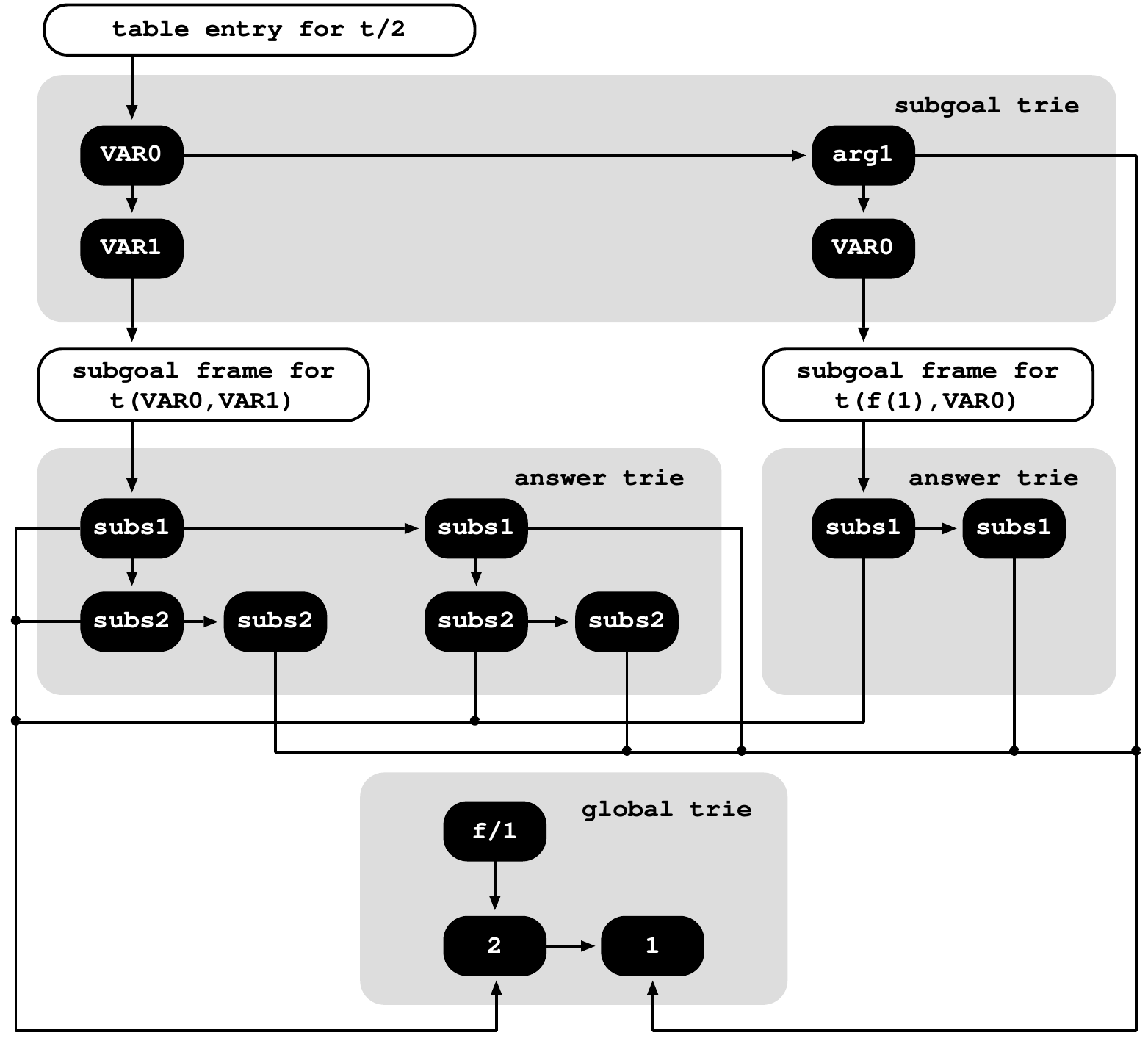}
\caption{GT-T's table space organization}
\label{fig_gt_t_design}
\end{figure}

Each path in a subgoal or answer trie is composed of a fixed number of
trie nodes, representing, in the subgoal trie, the number of arguments
for the corresponding tabled subgoal call, and, in the answer trie,
the number of substitution terms for the corresponding answer. More
specifically, for the subgoal tries, each node represents an argument
term $arg_i$ in which the node's token is used to store either
$arg_i$, if $arg_i$ is a \emph{simple term} (an atom, integer or
variable term), or the reference to the path's leaf node in the GT
representing $arg_i$, if $arg_i$ is a \emph{compound (non-simple)
  term}. Similarly for the answer tries, each node represents a
substitution term $subs_i$, where the node's token stores either
$subs_i$, if $subs_i$ is a simple term, or the reference to the path's
leaf node in the GT representing $subs_i$, if $subs_i$ is a compound
term. Figure~\ref{fig_gt_t_design} uses the same example from
Fig.~\ref{fig_original_design} to illustrate how the GT-T design
works.

Initially, the subgoal trie and the GT are empty. Then, the subgoal
\texttt{t(f(1),Y)} is called and the argument compound term
\texttt{f(1)} (represented by the tokens \texttt{f/1} and \texttt{1})
is first inserted in the GT. The two argument terms are then
represented in the subgoal trie (nodes \texttt{arg1} and
\texttt{VAR0}), where the node's token for \texttt{arg1} stores the
reference to the leaf node of the corresponding term representation
inserted in the GT. For the second subgoal call \texttt{t(X,Y)}, the
argument terms \texttt{VAR0} and \texttt{VAR1}, representing
respectively \texttt{X} and \texttt{Y}, are both simple terms and thus
we simply insert two nodes in the subgoal trie to represent them.

When processing answers, the procedure is similar to the one described
above for the subgoal calls. For each substitution compound term
\texttt{f(1)} and \texttt{f(2)}, we also insert first its
representation in the GT and then we insert a node in the
corresponding answer trie (nodes labeled \texttt{subs1} and
\texttt{subs2} in Fig.~\ref{fig_gt_t_design}) storing the reference to
its path in the GT. As \texttt{f(1)} was inserted in the GT at the
time of the first subgoal call, we only need to insert \texttt{f(2)}
(represented by the nodes \texttt{f/1} and \texttt{2}), meaning that
in fact we only need to insert the token \texttt{2} in the GT, in
order to represent the full set of answers. So, we are maximizing the
sharing of common terms appearing at different arguments or
substitution positions. For this particular example, the result is a
very compact representation of the GT, as most subgoal calls and/or
answers share the same term representations.

On completion of a subgoal, a strategy exists that avoids loading
answers from the answer tries using bottom-up unification, performing
instead what is called a \emph{completed table
  optimization}~\cite{RamakrishnanIV-99}. This optimization implements
answer recovery by top-down traversing the completed answer trie and
by executing specific WAM-like instructions from the answer trie
nodes. In the GT-T design, the difference caused by the existence of
the GT is a new set of WAM-like instructions that, instead of working
at the level of atoms/terms/functors/lists as in the original
design~\cite{RamakrishnanIV-99}, work at the level of the substitution
terms. Consider, for example, the loading of four answers for the call
\texttt{t(X,Y)}. One has two choices for variable \texttt{X} and, to
each \texttt{X}, we have two choices for variable \texttt{Y}. In the
GT-T design, the answer trie nodes representing the choices for
\texttt{X} and for \texttt{Y} (nodes \texttt{subs1} and \texttt{subs2}
respectively) are compiled with a WAM-like sequence of trie
instructions, such as \texttt{try\_subs\_compound} (for first choices)
and \texttt{trust\_subs\_compound} (for second/last choices). GT-T's
compiled tries also include a \texttt{retry\_subs\_compound}
instruction (for intermediate choices), a \texttt{do\_subs\_compound}
instruction (for single choices) and similar variants for simple
(non-compound) terms: \texttt{do\_subs\_simple},
\texttt{try\_subs\_simple}, \texttt{retry\_subs\_simple} and
\texttt{trust\_subs\_simple}.

Regarding space reclamation, GT-T uses the child field of the leaf
nodes (that is always NULL for a leaf node in the GT) to count the
number of references to the path it represents. This feature is of
utmost importance for the deletion process of a path in the GT, which
can only be performed when there is no reference to it, this is true
when the leaf node's child field reaches zero.


\subsection{Global Trie for Subterms}

The GT-ST was designed taking into account the use of tabling in
problems where redundant data occurs more commonly. The GT-ST design
maintains most of the GT-T features, but tries to optimize GT's memory
usage by representing compound subterms in term arguments as unique
entries in the GT. Therefore, we maximize the sharing of the tabled
data that is structurally equal at a \emph{second level}, by avoiding
the representation of equal compound subterms, and thus preventing
situations where the representation of those subterms occur more than
once.

Although GT-ST uses the same GT-T's tree structure for implementing
the GT, every different path in the GT can now represent a complete
term or a subterm of another term, but still being an unique
term. Consider, for example the insertion of the term \texttt{f(g(1))}
in the GT. After storing the node representing functor \texttt{f/1},
the process is suspended and the subterm \texttt{g(1)} is inserted as
an individual term in the GT. After the complete insertion of subterm
\texttt{g(1)} in the GT, the insertion of the main term is resumed by
storing a node referencing the \texttt{g(1)} representation in the GT,
i.e., by storing a node referencing the leaf node that represents
\texttt{g(1)} in the GT.

\begin{figure}[!ht]
\centering
\includegraphics[width=11cm]{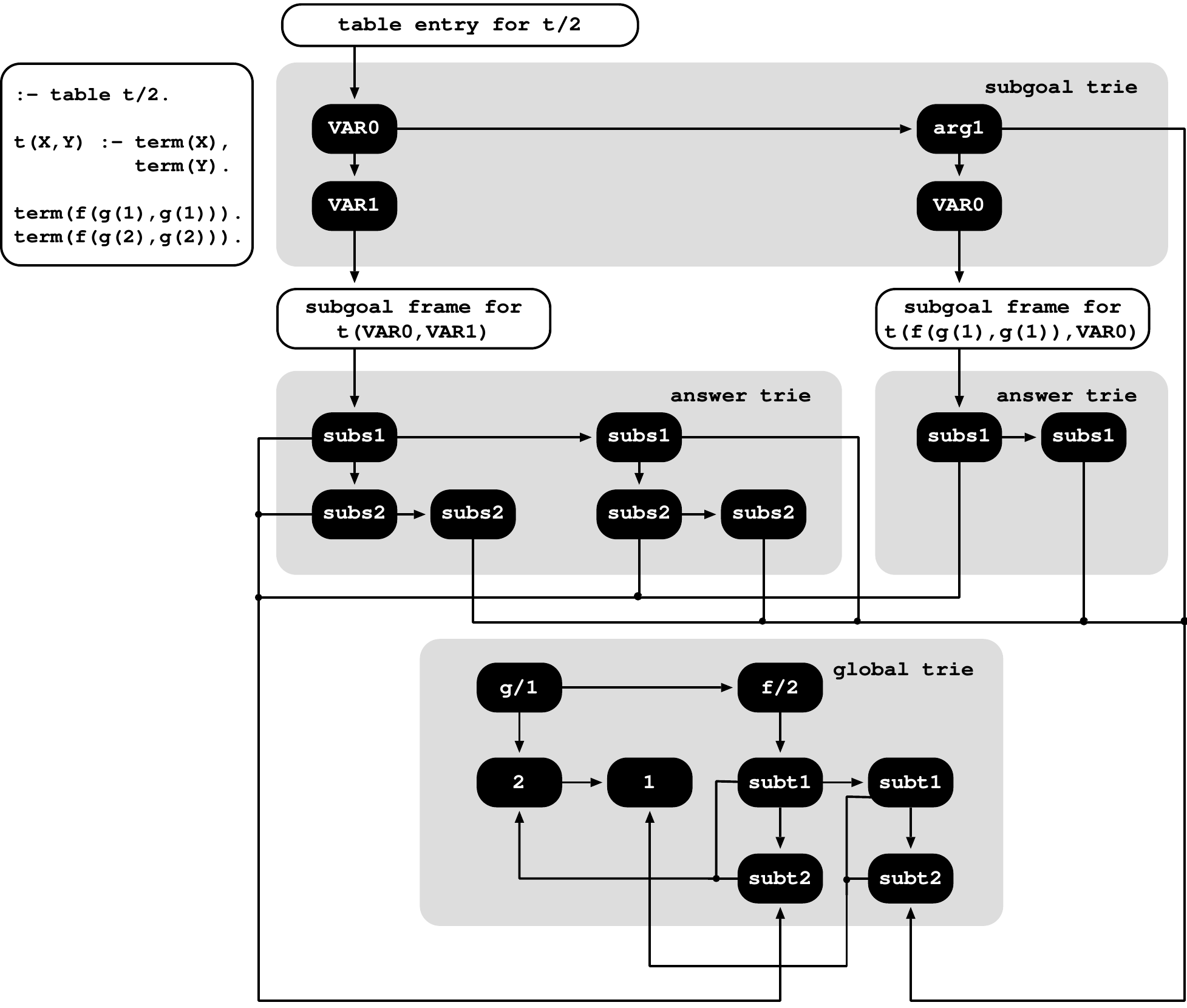}
\caption{GT-ST's table space organization}
\label{fig_gt_st_design}
\end{figure}

Despite these structural differences in the GT design, all the
remaining data structures remain unaltered. In particular, the GT-T's
structure for the subgoal and answer tries, where each path is
composed by a fixed number of nodes representing, respectively, the
number of arguments for table subgoal calls and the number of
substitution terms for tabled answers, is used without
changes. Moreover, features regarding the subgoal frame structure used
to maintain the chronological order of answers and to implement answer
recovery, also remain unchanged. Figure~\ref{fig_gt_st_design} shows
an example of how the GT-ST design works by illustrating the resulting
data structures for a tabled program with compound subterms.

Initially, the subgoal trie and the GT are empty. Then, a first
subgoal call occurs, \texttt{t(f(g(1),g(1)),Y)}, and the two argument
terms for the call are inserted in the subgoal trie with the compound
term being first inserted in the GT. Regarding the insertion of the
compound term \texttt{f(g(1),g(1))} in the GT, we next emphasize the
differences between the GT-ST and the GT-T designs.

At first, a node is inserted to represent the functor \texttt{f/2},
but then the insertion of the first subterm \texttt{g(1)} is
suspended, since \texttt{g(1)} is a compound term. The compound term
\texttt{g(1)} is then inserted as a distinct term in the GT and two
nodes, for functor \texttt{g/1} and integer \texttt{1}, are then
inserted in the GT with the node for functor \texttt{g/1} being a
sibling of the already stored node for functor \texttt{f/2}. After
storing \texttt{g(1)} in the GT, the insertion of the main term
\texttt{f(g(1),g(1))} is resumed and a new node, referencing the leaf
node of \texttt{g(1)}, is inserted as a child node of the node for
functor \texttt{f/2}. The construction of the main term then continues
by applying an analogous procedure to its second argument,
\texttt{g(1)}. However, the term \texttt{g(1)} is already stored in
the GT, therefore it is only required the insertion of a new node
referencing again the leaf node of \texttt{g(1)}.

As for the GT-T design, for the second subgoal call \texttt{t(X,Y)},
we do not interact with the GT. Both arguments are simple terms and
thus we simply insert two nodes, \texttt{VAR0} and \texttt{VAR1}, in
the subgoal trie to represent them.

The procedure used for processing answers is similar to the one just
described for the subgoal calls. For each substitution compound term,
we first insert the term in the GT and then we insert a node in the
corresponding answer trie storing the reference to its path in the GT
(nodes labeled \texttt{subs1} and \texttt{subs2} in
Fig.~\ref{fig_gt_st_design}). The complete set of answers for both
subgoal calls is formed by the substitution terms
\texttt{f(g(1),g(1))} and \texttt{f(g(2),g(2))}. Thus, as
\texttt{f(g(1),g(1))} was already inserted in the GT when storing the
first subgoal call, only \texttt{f(g(2),g(2))} needs to be stored in
order to represent the whole set of answers. As we are maximizing the
sharing of common subterms appearing at different argument or
substitution positions, for this particular example, this results in a
very compact representation of the GT.

Regarding the completed table optimization and space reclamation, the
GT-ST design implements the same GT-T's mechanisms described
previously. In particular, for space reclamation, the use of the child
field of the leaf nodes (that is always NULL for a leaf node in the
GT) to count the number of references to the path it represents can be
used as before for subterm's counting.


\section{Implementation Details}

\begin{figure}[!ht]
\centering
\includegraphics[width=11.5cm]{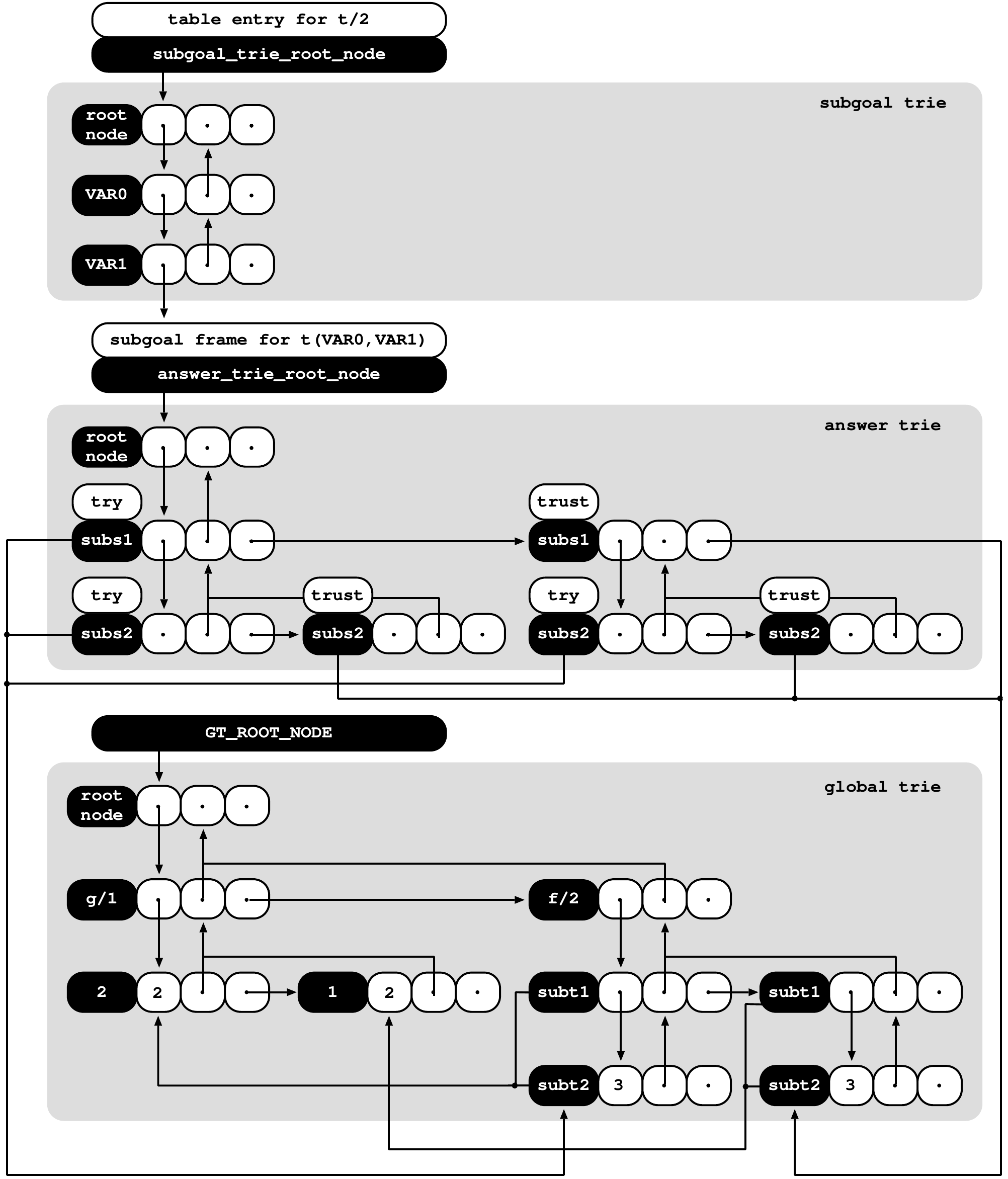}
\caption{Implementation details for the GT-ST's table space organization}
\label{fig_gt_st_implementation}
\end{figure}

We then describe the data structures and algorithms for GT-ST's table
space design. Figure~\ref{fig_gt_st_implementation} shows in more
detail the table organization previously presented in
Fig.~\ref{fig_gt_st_design} for the subgoal call \texttt{t(X,Y)}.

Internally, all tries are represented by a top \emph{root node},
acting as the entry point for the corresponding subgoal, answer or
global trie data structure. For the subgoal tries, the root node is
stored in the corresponding table entry's
\texttt{subgoal\_trie\_root\_node} data field. For the answer tries,
the root node is stored in the corresponding subgoal frame's
\texttt{answer\_trie\_root\_node} data field. For the GT, the root
node is stored in the \texttt{GT\_ROOT\_NODE} global variable.

Regarding trie nodes, they are internally implemented as 4-field data
structures. The first field (\texttt{token}) stores the token for the
node and the second (\texttt{child}), third (\texttt{parent}) and
fourth (\texttt{sibling}) fields store pointers, respectively, to the
first child node, to the parent node, and to the next sibling
node. Remember that for the GT, the leaf node's
\texttt{child} field is used to count the number of references to the
path it represents. For the answer tries, an additional field
(\texttt{code}) is used to support compiled tries.

Traversing a trie to check/insert for new calls or for new answers is
implemented by repeatedly invoking a
\texttt{trie\_token\_check\_insert()} procedure for each token that
represents the call/answer being checked. Given a trie node \texttt{n}
and a token \texttt{t}, the \texttt{trie\_token\_check\_insert()}
procedure returns the child node of \texttt{n} that represents the
given token \texttt{t}. Initially, the procedure traverses
sequentially the list of sibling nodes checking for one representing
the given token \texttt{t}. If no such node is found, a new trie node
is initialized and inserted in the beginning of the list. Searching
through a list of sibling nodes could be too expensive if we have
hundreds of siblings. A threshold value (8 in our implementation)
controls whether to dynamically index the nodes through a hash table,
hence providing direct node access and optimizing search. Further hash
collisions are reduced by dynamically expanding the hash tables. For
simplicity of presentation, in what follows, we omit the hashing
mechanism.

When inserting terms in the table space we need to distinguish two
situations: (i) inserting tabled calls in a subgoal trie structure;
and (ii) inserting answers in a particular answer trie
structure. These two situations are handled by the
\texttt{trie\_subgoal\_check\_insert()} and
\texttt{trie\_answer\_check\_insert()} procedures, respectively. The
pseudo-code for the \texttt{trie\_subgoal\_check\_insert()} procedure
is shown in Fig.~\ref{fig_trie_subgoal_check_insert}. The
\texttt{trie\_answer\_check\_insert()} procedure works similarly.

\begin{figure}[!ht]
{\verbatimproperties
\begin{verbatim}
trie_subgoal_check_insert(TABLE_ENTRY te, SUBGOAL_CALL call) {
  sg_node = te->subgoal_trie_root_node
  arity = get_arity(call)
  for (i = 1; i <= arity; i++) {
    t = get_argument_term(call, i)
    if (is_simple_term(t))
      sg_node = trie_token_check_insert(sg_node, t)
    else {                                        // t is a compound term
      gt_node = trie_term_check_insert(GT_ROOT_NODE, t)
      sg_node = trie_token_check_insert(sg_node, gt_node)
    }
  }
  return sg_node
}
\end{verbatim}}
\caption{Pseudo-code for the \texttt{trie\_subgoal\_check\_insert()} procedure}
\label{fig_trie_subgoal_check_insert}
\end{figure}

For each argument term \texttt{t} of the given subgoal call, the
procedure first checks if it is a simple term. If so, \texttt{t} is
inserted in the current subgoal trie. Otherwise, \texttt{t} is first
inserted in the GT and, then, it uses the reference to the leaf node
representing \texttt{t} in the GT (\texttt{gt\_node} in
Fig.~\ref{fig_trie_subgoal_check_insert}) as the token to be inserted
in the current subgoal trie.

The main difference to the previous GT-T design relies in the
insertion of terms in the GT, and for that we have changed the
\texttt{trie\_term\_check\_insert()} procedure in such a way that when
a compound term has compound subterms as arguments, the procedure
calls itself. Figure~\ref{fig_trie_term_check_insert} shows the
pseudo-code for the changes made to the
\texttt{trie\_term\_check\_insert()} procedure in order to support the
new GT-ST design.

As we can see in Fig.~\ref{fig_trie_subgoal_check_insert}, the initial
call to the \texttt{trie\_term\_check\_insert()} procedure is always
made with \texttt{GT\_ROOT\_NODE} as the first argument and with a
compound term as the second argument (respectively, arguments
\texttt{gt\_node} and \texttt{t} in
Fig.~\ref{fig_trie_term_check_insert}).

\begin{figure}[!ht]
{\verbatimproperties
\begin{verbatim}
trie_term_check_insert(TRIE_NODE gt_node, TERM t) {
  if (is_simple_term(t))
    gt_node = trie_token_check_insert(gt_node, t)
  else {                                          // t is a compound term
    if (gt_node == GT_ROOT_NODE) {
      name = get_name(t)
      arity = get_arity(t)
      gt_node = trie_token_check_insert(gt_node, name)
      for (i = 1; i <= arity; i++) {
        sub_t = get_argument_term(t, i)
        gt_node = trie_term_check_insert(gt_node, sub_t)
      }
    } else {                // t is a compound subterm of a compound term
      sub_gt_node = trie_term_check_insert(GT_ROOT_NODE, t)
      gt_node = trie_token_check_insert(gt_node, sub_gt_node)
    }
  }
  return gt_node
}
\end{verbatim}}
\caption{Pseudo-code for the \texttt{trie\_term\_check\_insert()} procedure}
\label{fig_trie_term_check_insert}
\end{figure}

Initially, the \texttt{trie\_term\_check\_insert()} procedure checks
if \texttt{t} is a simple term (always false for the initial call)
and, if so, \texttt{t} is simply inserted in the GT as a child node of
the given \texttt{gt\_node}. Otherwise, \texttt{t} is a compound term
and two situations can occur: (i) if \texttt{gt\_node} is
\texttt{GT\_ROOT\_NODE}, then the term's name is inserted in the GT
and, for each subterm of \texttt{t}, the procedure is invoked
recursively; (ii) if \texttt{gt\_node} is not \texttt{GT\_ROOT\_NODE},
which means that \texttt{t} is a compound subterm of a compound term,
the procedure calls itself with \texttt{GT\_ROOT\_NODE} as the first
argument. By doing that, \texttt{t} is inserted as a unique term in
the GT. When the procedure returns, the reference
\texttt{sub\_gt\_node} to the leaf node of the subterm's path
representation of \texttt{t} in the GT is inserted as a child node of
the given \texttt{gt\_node}.

Regarding the traversal of the answer tries to consume answers, the
GT-ST design follows the same implementation as in the GT-T design,
and, in particular, for compound terms, it uses a
\texttt{trie\_term\_load()} procedure to load, from the GT back to the
Prolog engine, the substitution term given by the reference stored in
the corresponding token field. The main difference to the previous
GT-T design is in the cases of subterm references in the GT, where the
\texttt{trie\_term\_load()} procedure calls itself to first load the
subterm reference from the GT.


\section{Experimental Results}

We next present some experimental results comparing YapTab with and
without support for the GT-T and GT-ST designs. The environment for
our experiments was a PC with a 2.66 GHz Intel(R) Core(TM) 2 Quad CPU
and 4 GBytes of memory running the Linux kernel 2.6.24 with YapTab
6.2.0.

To put the performance results in perspective and have a well-defined
starting point comparing the GT-T and GT-ST approaches, first we have
defined a tabled predicate \texttt{t/5} that simply stores in the
table space terms defined by \texttt{term/1} facts, and then we used a
top query goal \texttt{test/0} to recursively call \texttt{t/5} with
all combinations of one and two free variables in the arguments. An
example of such code for functor terms of arity 1 (1,000 terms in
total) is shown next.

{\verbatimproperties
\begin{verbatim}
   :- table t/5.
   t(A,B,C,D,E) :- term(A), term(B), term(C), term(D), term(E).
   
   test :- t(A,f(1),f(1),f(1),f(1)), fail.       term(f(1)).
   ...                                           term(f(2)).
   test :- t(f(1),f(1),f(1),f(1),A), fail.       term(f(3)).
   test :- t(A,B,f(1),f(1),f(1)), fail.          ...
   ...                                           term(f(998)).
   test :- t(f(1),f(1),f(1),A,B), fail.          term(f(999)).
   test.                                         term(f(1000)).
\end{verbatim}}

We experimented the \texttt{test/0} predicate with 9 different kinds
of 1,000 \texttt{term/1} facts: integers, atoms, functor (with arity
1, 2, 4 and 6) and list (with length 1, 2 and 4)
terms. Table~\ref{tab_terms} shows the table memory usage (column {\bf
  \emph{Mem}}), in MBytes, and the execution times, in milliseconds,
to store (column {\bf \emph{Str}}) the tables (first execution) and to
load from the tables (second execution) the complete set of answers
without (column {\bf \emph{Ld}}) and with (column {\bf \emph{Cmp}})
compiled tries for YapTab's original table design (column {\bf
  \emph{YapTab}}) and for the GT-T (column {\bf \emph{GT-T/YapTab}})
and GT-ST (column {\bf \emph{GT-ST/YapTab}}) designs. For GT-T and
GT-ST, we only show the ratios over YapTab's original table
design. The execution times are the average of five runs.

\begin{table}[!ht]
\tableproperties
\begin{tabular}{l|rrrr|cccc|cccc}
\hline\hline
\multicolumn{1}{c|}{\multirow{2}{*}{\bf \emph{Terms}}}
& \multicolumn{4}{c|}{\bf \emph{YapTab}}
& \multicolumn{4}{c|}{\bf \emph{GT-T/YapTab}}
& \multicolumn{4}{c}{\bf \emph{GT-ST/YapTab}}
\\
& \multicolumn{1}{c}{\bf \emph{Mem}}
& \multicolumn{1}{c}{\bf \emph{Str}}
& \multicolumn{1}{c}{\bf \emph{Ld}}
& \multicolumn{1}{c|}{\bf \emph{Cmp}}
& {\bf \emph{Mem}}
& {\bf \emph{Str}}
& {\bf \emph{Ld}}
& {\bf \emph{Cmp}}
& {\bf \emph{Mem}}
& {\bf \emph{Str}}
& {\bf \emph{Ld}}
& {\bf \emph{Cmp}}
\\
\hline
\textbf{1,000 ints}   &   191 & 1,270 &   345 &   344 &         1.00  &         1.05  &         1.00  &         1.00  &         1.00  &         1.09  &         1.11  &         1.07 \\
\textbf{1,000 atoms}  &   191 & 1,423 &   343 &   406 &         1.00  &         1.04  &         1.01  &         1.02  &         1.00  &         1.04  &         1.03  &         1.08 \\
\textbf{1,000 f/1}    &   191 & 1,680 &   542 &   361 &         1.00  &         1.32  &         1.16  &         2.10  &         1.00  &         1.34  &         1.17  &         2.13 \\
\textbf{1,000 f/2}    &   382 & 2,295 &   657 &   450 & \textbf{0.50} &         1.10  &         1.14  &         1.84  & \textbf{0.50} &         1.06  &         1.11  &         1.88 \\
\textbf{1,000 f/4}    &   764 & 3,843 &   973 &   631 & \textbf{0.25} & \textbf{0.81} & \textbf{0.98} &         1.44  & \textbf{0.25} & \textbf{0.78} &         1.04  &         1.53 \\
\textbf{1,000 f/6}    & 1,146 & 5,181 & 1,514 &   798 & \textbf{0.17} & \textbf{0.72} & \textbf{0.72} &         1.38  & \textbf{0.17} & \textbf{0.66} & \textbf{0.71} &         1.36 \\
\textbf{1,000 [ ]/1}  &   382 & 2,215 &   507 &   466 & \textbf{0.50} &         1.08  &         1.05  &         1.61  & \textbf{0.50} &         1.10  &         1.02  &         1.58 \\
\textbf{1,000 [ ]/2}  &   764 & 3,832 &   818 &   604 & \textbf{0.25} & \textbf{0.80} & \textbf{0.94} &         1.38  & \textbf{0.25} &         1.00  &         1.05  &         1.48 \\
\textbf{1,000 [ ]/4}  & 1,528 & 6,566 & 1,841 & 1,066 & \textbf{0.13} & \textbf{0.63} & \textbf{0.54} & \textbf{0.96} & \textbf{0.13} & \textbf{0.89} & \textbf{0.66} &         1.14 \\
\hline                                                                                                                                                       
\multicolumn{5}{l|}{\bf \emph{Average}}              & \textbf{0.53} & \textbf{0.95} & \textbf{0.95} & \textbf{1.42} & \textbf{0.53} & \textbf{0.99} & \textbf{0.99} & \textbf{1.47} \\
\hline\hline
\end{tabular}
\caption{Table memory usage (in MBytes) and store/load times (in
  milliseconds) comparing YapTab's original table design with the GT-T
  and GT-ST designs}
\label{tab_terms}
\end{table}

The results in Table~\ref{tab_terms} suggest that both GT designs are
a very good approach to reduce memory usage and that this reduction
increases proportionally to the length and redundancy of the terms
stored in the GT. In particular, for functor and list terms, the
results show an increasing and very significant reduction on memory
usage, for both GT-T and GT-ST approaches. The results for the special
cases of integer and atom terms are also very interesting as they show
that the cost of representing only simple terms in the respective
tries. Note that, although, integer and atom terms are only
represented in the respective tries, it is necessary to check for
these types of terms, in order to proceed with the respective
store/load algorithm.

Regarding execution time, the results suggest that, in general, GT-ST
spends more time in the store and load term procedures than GT-T. Such
behaviour can be easily explained by the fact that, the GT-ST's
storing and loading algorithms have more sub-cases to process in order
to support subterms. These results also seem to indicate that memory
reduction for small sized terms, generally comes at a price in storing
time (between 4\% and 32\% more for GT-T and between 4\% and 34\% more
for GT-ST in these experiments). The opposite occurs in the tests
where term's length are higher (between 19\% and 37\% less for GT-T
and 11\% and 34\% less for GT-ST).  Note that with GT-T and GT-ST
support, we pay the cost of navigating in two tries when
checking/storing/loading a term. Moreover, in some situations, the
cost of storing a new term in an empty/small trie can be less than the
cost of navigating in the GT, even when the term is already stored in
the GT. However, our results seem to suggest that this cost decreases
proportionally to the length and redundancy of the terms stored in the
GT. In particular, for functor and list terms, GT-T and GT-ST support
showed to outperform the original YapTab design when we increase the
length of the terms stored in the GT.

The results obtained for loading terms also show some gains without
compiled tries (around 5\% for GT-T and 1\% for GT-ST on average) but,
when using compiled tries the results show some significant costs on
execution time (around 42\% for GT-T and 47\% for GT-ST on
average). We believe that this cost is smaller for GT-T as a result of
having less sub-cases in the storing/loading algorithms. On the other
hand, we also believe that some cache behaviour effects, reduce the
costs on execution time, for both GT designs. As we need to navigate
in the GT for each substitution term, we kept accessing the
same GT nodes, thus reducing eventual cache misses. This
seems to be the reason why for list terms of length 4, GT-T
outperforms the original YapTab design, both without and with compiled
tries. Note that, for this particular case, both GT-T and GT-ST only
consumes 13\% of the memory used with the original YapTab design.

Next, we experimented with a new set of tests specially designed to
provide more expressive results regarding the comparison between the
GT-ST and the GT-T designs. In this tests, we have defined a tabled
predicate \texttt{t/1} that simply stores in the table space terms
defined by \texttt{term/1} facts and then we used a \texttt{test/0}
predicate to call \texttt{t/1} with a free variable. We experimented
the \texttt{test/0} predicate with 9 different sets of 500,000 term
facts of compound terms (with arity 1, 2, 3) where its arguments are
also compound subterms (with arity 1, 3, 5). An example of such code
for a functor term \texttt{f/2} with argument subterms \texttt{g/3}
(500,000 terms in total) is shown next.

{\verbatimproperties
\begin{verbatim}
   :- table t/1.                          test :- t(A), fail.
   t(A) :- term(A).                       test.           

   term(f(g(1,1,1), g(1,1,1))).
   term(f(g(2,2,2), g(2,2,2))).
   term(f(g(3,3,3), g(3,3,3))).
   ...
   term(f(g(499998,499998,499998), g(499998,499998,499998))).
   term(f(g(499999,499999,499999), g(499999,499999,499999))).
   term(f(g(500000,500000,500000), g(500000,500000,500000))).
\end{verbatim}}

As opposed to the previous experiments, here we just used one free
variable for the tabled predicate \texttt{t/1}. This difference is
necessary because, when we have more than one free variable and we
produce different combinations between those free variables, we are
raising the number of nodes represented in the local tries. More
precisely, different combinations of free variables raises the number
of answers and therefore the number of nodes in the local answer
tries.

Table~\ref{tab_subterms} shows the table memory usage (columns {\bf
  \emph{Memory}}) composed by two columns, one for total memory
(columns {\bf \emph{Total}}) and the other for GT's memory only
(columns {\bf \emph{GT}}), in MBytes, and the execution times, in
milliseconds, to store (columns {\bf \emph{Str}}) the tables (first
execution) and to load from the tables (second execution) the complete
set of answers without (columns {\bf \emph{Ld}}) and with (columns
{\bf \emph{Cmp}}) compiled tries using the GT-T table design (column
{\bf \emph{GT-T}}), and using the GT-ST design (column {\bf
  \emph{GT-ST/GT-T}}). For the values referring the GT-ST we only show
the ratios over the GT-T design. Since the main purpose of this second
set of experiments is to compare the differences between GT-T and
GT-ST, we did not include YapTab's original table design in these
experiments. The execution times are the average of five runs.

\begin{table}[!ht]
\tableproperties
\begin{tabular}{l|rrrrr|ccccc}
\hline\hline
& \multicolumn{5}{c|}{\bf \emph{GT-T}}
& \multicolumn{5}{c}{\bf \emph{GT-ST/GT-T}} 
\\
\multicolumn{1}{c|}{\bf \emph{Terms}}
& \multicolumn{2}{c}{\bf \emph{Memory}} & & & 
& \multicolumn{2}{c}{\bf \emph{Memory}} & & &
\\
& \multicolumn{1}{c}{\bf \emph{Total}}
& \multicolumn{1}{c}{\bf \emph{GT}}
& \multicolumn{1}{c}{\bf \emph{Str}}
& \multicolumn{1}{c}{\bf \emph{Ld}}
& \multicolumn{1}{c|}{\bf \emph{Cmp}}
& {\bf \emph{Total}}
& {\bf \emph{GT}}
& {\bf \emph{Str}}
& {\bf \emph{Ld}}
& {\bf \emph{Cmp}}
\\
\hline
\textbf{f/1} & & & & & \\
\textbf{500,000 g/1} &  17.17 &  7.63 & 126 &  28 &  51 &         1.44  &         2.00  &         1.55  &         1.14  &         1.00  \\
\textbf{500,000 g/3} &  32.43 & 22.89 & 198 &  34 &  61 &         1.24  &         1.33  &         3.29  &         1.12  &         1.25  \\
\textbf{500,000 g/5} &  47.68 & 38.15 & 293 &  47 &  83 &         1.16  &         1.20  &         1.46  &         1.00  & \textbf{0.99} \\
\hline
\textbf{f/2} & & & & & \\
\textbf{500,000 g/1} &  32.43 & 22.89 & 203 &  38 &  71 &         1.00  &         1.00  &         1.28  &         1.13  &         1.09  \\
\textbf{500,000 g/3} &  62.94 & 53.41 &  45 &  60 & 103 & \textbf{0.76} & \textbf{0.71} &         1.18  & \textbf{0.84} & \textbf{0.95} \\
\textbf{500,000 g/5} &  93.46 & 83.92 & 438 & 111 & 146 & \textbf{0.67} & \textbf{0.64} &         1.10  & \textbf{0.67} & \textbf{0.80} \\
\hline
\textbf{f/3} & & & & & \\
\textbf{500,000 g/1} &  47.68 & 38.15 & 296 &  50 &  89 & \textbf{0.84} & \textbf{0.80} &         2.87  &         1.02  &         1.03  \\
\textbf{500,000 g/3} &  93.46 & 83.92 & 616 & 142 & 164 & \textbf{0.59} & \textbf{0.55} &         1.25  & \textbf{0.80} & \textbf{0.85} \\
\textbf{500,000 g/5} & 139.24 & 129.7 & 832 & 197 & 224 & \textbf{0.51} & \textbf{0.47} & \textbf{0.96} & \textbf{0.67} & \textbf{0.74} \\
\hline
\multicolumn{6}{l|}{\bf \emph{Average}}                 & \textbf{0.96} & \textbf{0.97} & \textbf{0.93} & \textbf{0.97} & \textbf{0.91} \\
\hline\hline
\end{tabular}
\caption{Table memory usage (in MBytes) and store/load times (in
  seconds) comparing the GT-T and GT-ST designs for subterm
  representation}
\label{tab_subterms}
\end{table}

The results in Table~\ref{tab_subterms} suggest that GT-ST support has
potential to outperform GT-T's design with significant reductions on
memory usage and execution time for programs with increasing
redundancy on compound subterms. However, the results also show that,
for some base cases, the storing process can be a very expensive
procedure when compared with GT-T's design.

In general, the results suggest three different situations. For
\texttt{f/1} terms, the costs for GT-ST are globally higher. This
happens because GT-ST needs to store one extra node for every distinct
subterm representation and there is no redundancy in the \texttt{f/1}
subterms. However, the results show that the memory and execution
costs can be reduced when the subterm's arity increases from
\texttt{g/1} to \texttt{g/5}. This occurs because the cost of the
extra node for each subterm became diluted in the number of nodes
represented in the GT.

For \texttt{f/2} terms, a particular situation occurs for the case of
\texttt{g/1} subterms, where the memory spent is the same for both
designs. This happens because the extra node used by GT-ST, to
represent the reference to the subterm representation, is balanced by
the arity of the functor term \texttt{f/2}. From this point on, for
the remaining \texttt{f/2} terms and all the \texttt{f/3} terms, the
GT-ST always outperforms the GT-T, not only for the system's memory,
but also for the execution times with and without compiled
tries. These results suggest that, at least for some applications,
GT-ST support has potential to achieve significant reductions on
memory usage and execution time when compared with GT-T's design.


\section{Conclusions}

We have presented a new design for the table space organization, named
\emph{Global Trie for Subterms (GT-ST)}, that extends the previous
\emph{Global Trie for Terms (GT-T)} design. The GT-ST design maintains
most of the GT-T features, but tries to optimize GT's memory usage by
avoiding the representation of equal compound subterms, thus
preventing situations where the representation of those subterms occur
more than once and maximizing the sharing of the tabled data that is
structurally equal at a second level.

Experiments results, using the YapTab tabling system, show that GT-ST
support has potential to achieve significant reductions on memory
usage and execution time for programs with increasing compound
subterms in term arguments, without compromising the execution time
for other programs.

Further work will include seeking real-world applications, that pose
many subgoal queries possibly with a large number of redundant
answers, thus allowing us to improve and expand the current
implementation. In particular, we intend to study how
alternative/complementary designs for the table space organization can
further reduce redundancy in term representation.


\section*{Acknowledgments}

This work has been partially supported by the FCT research projects
HORUS (PTDC/EIA-EIA/100897/2008) and LEAP (PTDC/EIA-CCO/112158/2009).


\bibliographystyle{splncs}
\bibliography{references}

\end{document}